\documentclass[twocolumn]{aastex61}

\def\Mdot{\hbox{$\dot {M}$}}

\def\Rsun{\hbox{\it R$_\odot$}}

\def\Rstar{\hbox{\it R$_*$}}
\def\Rt{\hbox{R$_{\rm T}$}}
\def\Lsun{\hbox{\it L$_\odot$}}
\def\Lstar{\hbox{\it L$_*$}}
\def\Tstar{\hbox{\it T$_*$}}

\def\Msunyr{\hbox{\it M$_\odot\,$yr$^{-1}$}}

\def\Vinf{\hbox{$v_\infty$}}
\def\kms{\hbox{km$\,$s$^{-1}$}}

\def\simgr{\mathrel{\hbox{\rlap{\hbox{\lower4pt\hbox{$\sim$}}}\hbox{$>$}}}}

\shortauthors{Najarro et al.}
\shorttitle{Emission Lines in the IR Quintuplet Stars}

\begin{document}

\title{Emission Lines in the Near-infrared Spectra \\
of the Infrared Quintuplet 
Stars in the Galactic Center}

\author{F. Najarro}
\affiliation{Departamento de Astrof\'{\i}sica, Centro de Astrobiolog\'{\i}a (CSIC-INTA), Ctra. Torrej\'on a 
Ajalvir km 4, 28850 Torrej\'on de Ardoz, Spain }
\author{T. R. Geballe}
\affiliation{Gemini Observatory, 670 N. A'ohoku Place, Hilo, HI 96720}
\author{D. F. Figer}
\affiliation{Center for Detectors, Rochester Institute of Technology, 74 Lomb Memorial Drive,
Rochester, NY 14623, USA}
\author{D. de la Fuente}
\affiliation{Instituto de Astronom\'{\i}a, Unidad Acad\'emica en Ensenada,
Universidad Nacional Aut\'onoma de M\'exico, Ensenada 22860, M\'exico}

\begin{abstract}

We report the detection of a number of emission lines in the 1.0--2.4~$\mu$m spectra of four of the five bright infrared dust-embedded stars
at the center of the Galactic center's Quintuplet Cluster. Spectroscopy of the central stars of these objects is hampered not only by the large
interstellar extinction that obscures all objects in the Galactic center, but also by the large amounts of warm circumstellar dust surrounding 
each of the five. The pinwheel morphologies of the dust observed previously around two of
them are indicative of  Wolf-Rayet colliding wind binaries; however, infrared spectra of each of the five have until now revealed only dust continua steeply rising to long wavelengths and absorption lines and bands from interstellar gas and dust. The emission lines detected, from ionized carbon and from helium, are broad and confirm that the objects are dusty late-type carbon Wolf-Rayet stars. 

\end{abstract}


\vfill\eject

\section{Introduction}

Most stars in the central few tens of parsecs of the Galaxy are obscured from view by $\sim$30 visual 
magnitudes of extinction.
This is the case for almost all of the stars within the three massive stellar clusters there
\citep{fig04}. One of the clusters, the Quintuplet, is known for its extraordinary collection of five infrared-bright stars for which the cluster
was named \citep{nag90,oku90,gla90}. While hundreds of other massive stars in the cluster have been classified, the natures of these five objects,
the so-called Quintuplet Proper Members (QPMs) have remained mysterious \citep{fig99}. Light from them is not only obscured by the extinction to
the Galactic Center (GC), but also by their warm dusty cocoons, which add additional extinction and superimpose bright infrared continuum emission on whatever photospheric or wind emission might faintly leak out from within the cocoons. 

Previously published infrared spectra of the Quintuplet stars are  essentially featureless, apart from interstellar gas and solid phase absorptions, e.g., due to CO, H$_3^+$, CO$_2$, diffuse interstellar bands, and the 3.4-$\mu$m hydrocarbon and 9.7-$\mu$m silicate features, and fine structure and \ion{H}{1} recombination line emission from unrelated ionized gas \citep{oku90,fig99,chi00, mon01,oka05,lie09,geb11} along the line of sight. Because of this the QPMs have sometimes been used as telluric infrared spectroscopic standards for Galactic center sources with more ``interesting" spectra. 

Thus little has been known as to the natures of the QPMs.
\citet{fig99} and \citet{mon01} proposed that they are dusty late-type
Wolf-Rayet (WR) carbon stars (DWCLs). More recently, using speckle
imaging techniques in the $K$-band, \citet{tut06} resolved all five objects. They found that two of the sources have pinwheel-like nebulae, similar to those seen around two other Wolf-Rayet stars in the Galaxy, WR98a and WR104 \citep[][respectively]{mon99,tut99} and concluded that they are carbon-rich Wolf-Rayet stars (WCs).
The two pinwheel sources are GCS3-2 and GCS4, also known as Q2 and Q3,  respectively according to the commonly-used nomenclatures for these sources from \citet{nag90} and \citet{mon94}.

Late-type WC stars often are surrounded by dust. According to \citet{van01},
over half of WC9 types, and perhaps one-third of WC8 types, have dust emission.
\footnote{These fractions are confirmed in the uptated, more extended
{\it Galactic Wolf Rayet Catalogue} http://pacrowther.staff.shef.ac.uk/WRcat/index.php}
In such systems, the dust phenomenon can be persistent or episodic. In the most widely-accepted model, the pinwheel morphology is produced by warm dust that recedes in a radial direction from a binary system composed of a WR star and a less evolved secondary massive star, e.g.\ an O-star, whose winds collide. The ``Archimedean" spiral shape of the dust emission is a natural result of the rotating site of dust formation that is located on the opposite side of the secondary. Dust forms and propagates radially outward as the stars orbit each other. A similar pattern can be seen in water emitted from a garden hose as the hose is rotated around a central point.

Binarity is key to the pinwheel morphology \citep{tut99}. WR stars, dusty or not, are often in binary systems. The \citet{van01} catalog suggests that even for the current census of WR stars in the solar neighborhood the estimated binarity percentage of 40\% may be too low.  It is possible that all DWCLs are necessarily binary as a condition of dust production around them. 

\begin{figure} 
\centering
\begin{minipage}{4.0in}
{\includegraphics[scale=0.255,angle=0]{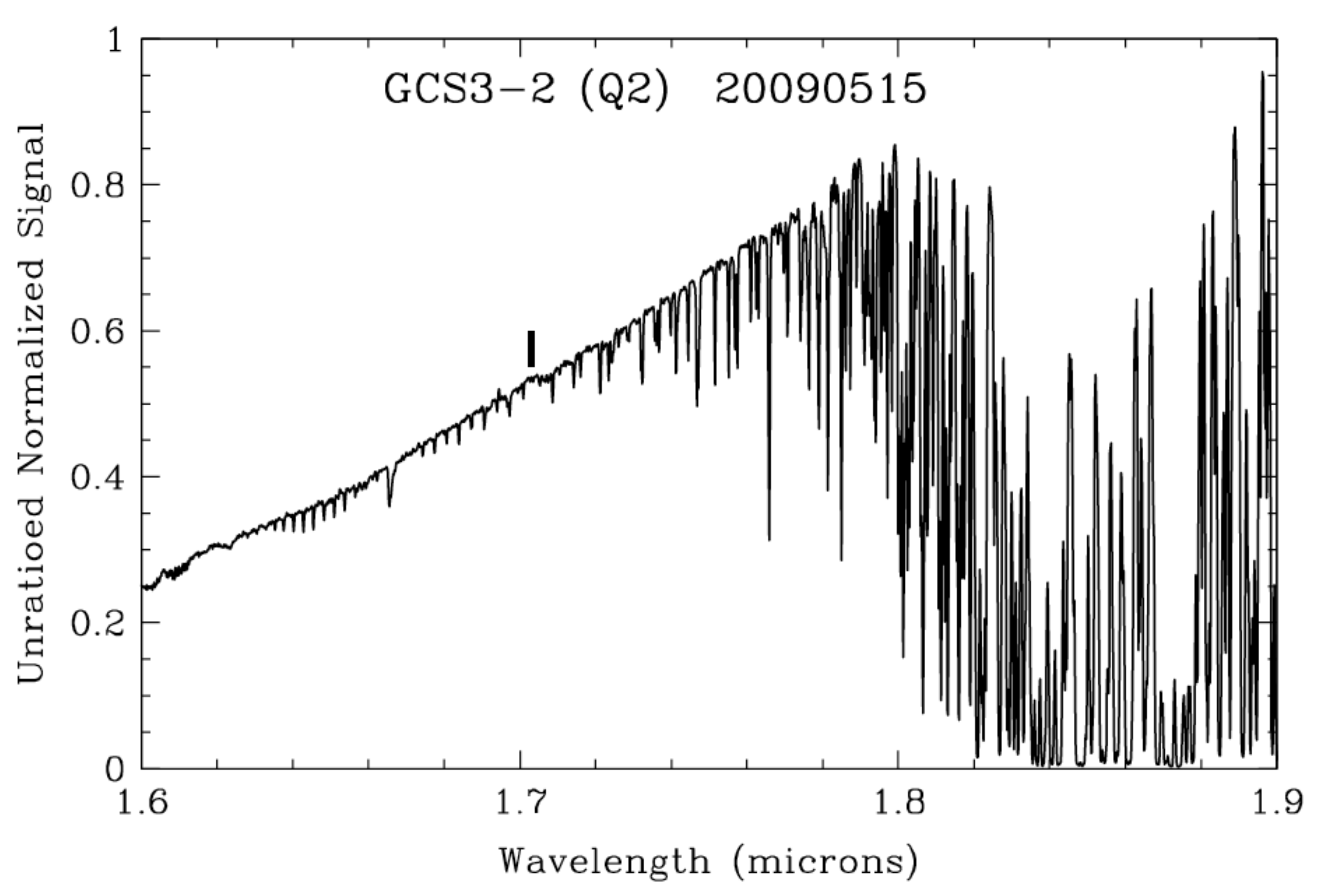}}
\end{minipage}
\caption{Unratioed $H$-band spectrum of GCS3-2 at $R$=5,000. A faint emission 
line, denoted by the vertical line, is present near 1.70~$\mu$m. Almost all of 
the absorption features are telluric.
\label{fig:q2raw}}
\end{figure}

On UT 2009 May 15, during the course of an unrelated observing
program, we observed GCS3-2 as a telluric standard and detected a weak
and broad emission line centered near 1.70~$\mu$m (see
Fig.~\ref{fig:q2raw}), presumably leaking out through the cocoon. The
detection led us to obtain spectra of the other IR Quintuplet stars in
the $H$-band and observe even shorter wavelengths. Although toward
shorter wavelengths the extinction due to interstellar dust and dust in the cocoons more severely attenuates any emerging spectrum from the central stars, the dust continua emitted by the cocoons weaken rapidly, as those wavelengths are on the Wien side of the spectral energy distributions (SEDs) of the dust. Because the severe dilution of line equivalent widths by dust emission is a major impediment to detecting lines from these embedded stars at longer infrared wavelengths, we considered that it might be possible to detect lines at shorter infrared wavelengths, despite the much weaker signal, by acquiring high-sensitivity spectra.

\section{Observations and Data Reduction}

$J$-, $H$-, and $K$-band spectra of the Quintuplet IR stars were obtained at the Frederick C. Gillett Gemini North Telescope at 
various times between 2009  May and 2012 June (programs GN-2009A-Q-25,
GN-2011A-Q-64, and GN-2012A-Q-61), using the facility instruments NIFS
(the Near-infrared Integral Field Spectrometer) and GNIRS (the Gemini
Near InfraRed Spectrograph) at resolving powers ($R$ =
$\lambda$/$\Delta \lambda$) ranging from 1,200 to 5,000. For NIFS the
telescope was alternately positioned so that the star was centered in the integral field and then that the field was observing nearby blank sky.  For GNIRS the standard stare/nod-along slit technique was employed.  Early A-type dwarfs served as telluric standards. 
An observing log is provided in  Table~1.

The NIFS spectra were obtained only in the $H$ band. The GNIRS spectra
were obtained in that instrument's cross-dispersed mode covering
0.9-2.5~$\mu$m. Because the flux densities of the Quintuplet stars increase rapidly at longer wavelengths, the long exposures required to maximize the sensitivity in the $J$ and $H$ bands resulted in the $K$-band spectra saturating that part of the detector array for GCS3-1, GCS3-4, and GCS4. Additional GNIRS spectra of those three stars were obtained using shorter exposure times so that their $K$-band spectra were not saturated. The only cross-dispersed spectrum of GCS3-2 used short exposure times and no signal was detected in the  0.9--1.4~$\mu$m region. 

Data reduction was standard, consisting of extracting spectra from subtracted pairs of nodded frames, spike removal, wavelength calibration utilizing both emission from arc lamps and telluric absorption lines in the astronomical spectra, wavelength-shifting to align the spectra of the Quintuplet stars with those of their telluric standards prior to ratioing them, and correcting for H I absorption lines in the spectra of the telluric standards. 

Because the reduced spectra rise steeply to longer wavelengths (see Fig.~\ref{fig:q2raw}), they are displayed here with their continua normalized. As a result of this, the spectra in Fig.~\ref{fig:q_j} appear increasingly noisy at the shorter wavelengths, where the continua are weakest and the emission lines are most heavily attenuated.

\begin{deluxetable}{lcccc}
\tablewidth{0pt}
\tablecaption{Observing Log}
\tablehead{\colhead{Object} & \colhead{Date} & \colhead{Band} & \colhead{Instrument} & \colhead{$R$}}
\startdata
GCS3-2 (Q2) & 20090515  & H  & NIFS & 5000 \\
~GCS3-2 (Q2) & 20100715  & H  & NIFS & 5000 \\
~GCS3-2 (Q2) & 20110429  & J,H,K & GNIRS & 1200 \\
~GCS3-1 (Q4) & 20110531  & J,H & GNIRS & 1200 \\
~GCS3-1 (Q4) & 20110607 & K & GNIRS & 1200 \\
~GCS3-4 (Q1) & 20110622 & K & GNIRS & 1200 \\
~GCS3-3 (Q9) & 20110703 & J,H & GNIRS & 1200 \\
~GCS3-4 (Q1) & 20110705  & J,H & GNIRS & 1200 \\
~GCS4 (Q3) & 20120604  & J,H,K & GNIRS & 4000 \\
\enddata
\label{tab:obs}
\end{deluxetable}

\section{Results}

\subsection{Detected lines}

\begin{figure} 
\centering
\begin{minipage}{4.0in}
{\includegraphics[scale=0.37,angle=0]{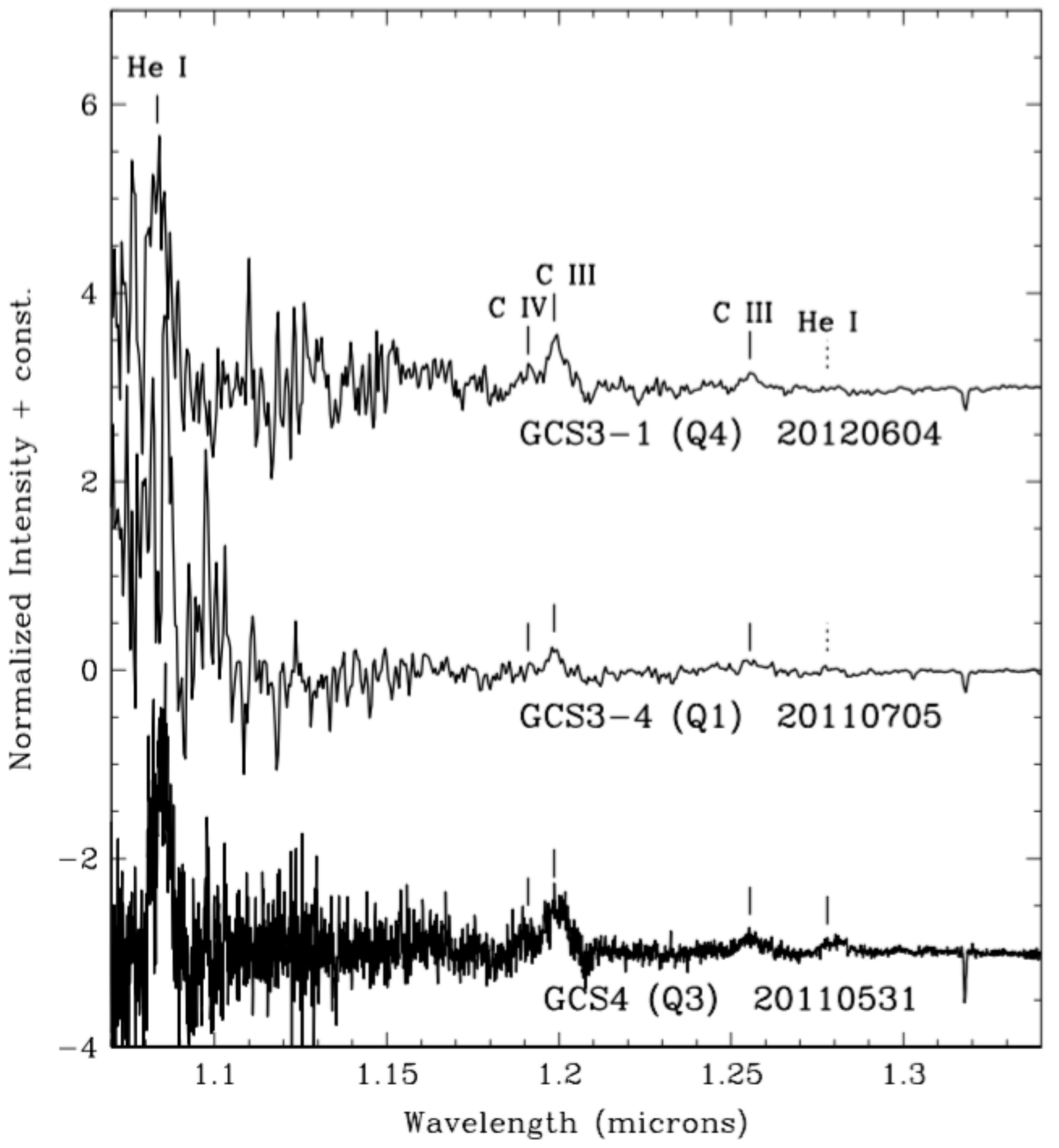}}
\end{minipage}
\caption{$J$--band continuum-normalized spectra
of GCS3-1 ($R$=1200), GCS3-4 ($R$=1200), and GCS4 ($R$=4000).  Wavelengths and identifications of clearly and marginally detected emission lines are shown by continuous and dashed lines, respectively. The absorption features at 1.318~$\mu$m is a diffuse interstellar band \citep{job90}. \label{fig:q_j}}
\end{figure}

\begin{figure} 
\centering
\begin{minipage}{4.0in}
{\includegraphics[scale=0.363,angle=0]{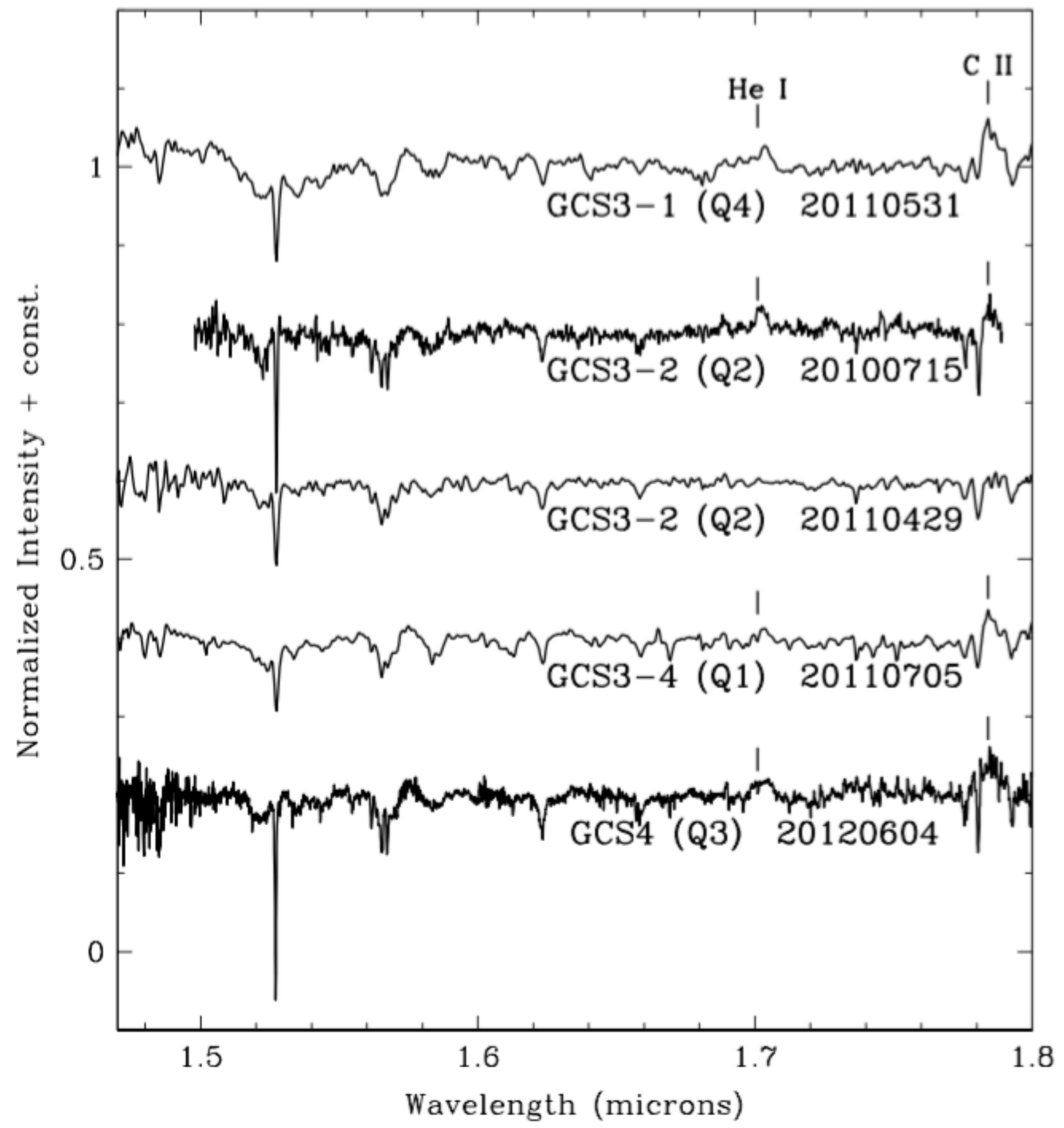}}
\end{minipage}
\caption{$H$-band continuum-normalized spectra of GCS3-1 ($R$=1200), GCS3-2 (i2010: $R$=4000, 2011: $R$=1200), GCS3-4 ($R$=1200), and GCS4 ($R$=4000). Wavelengths and identifications of clearly detected emission lines are indicated. Absorption features and complexes are diffuse interstellar bands \citep{geb11}. 
\label{fig:q_h}}
\end{figure}

\begin{figure} 
\centering
\begin{minipage}{4.0in}
{\includegraphics[scale=0.275,angle=0]{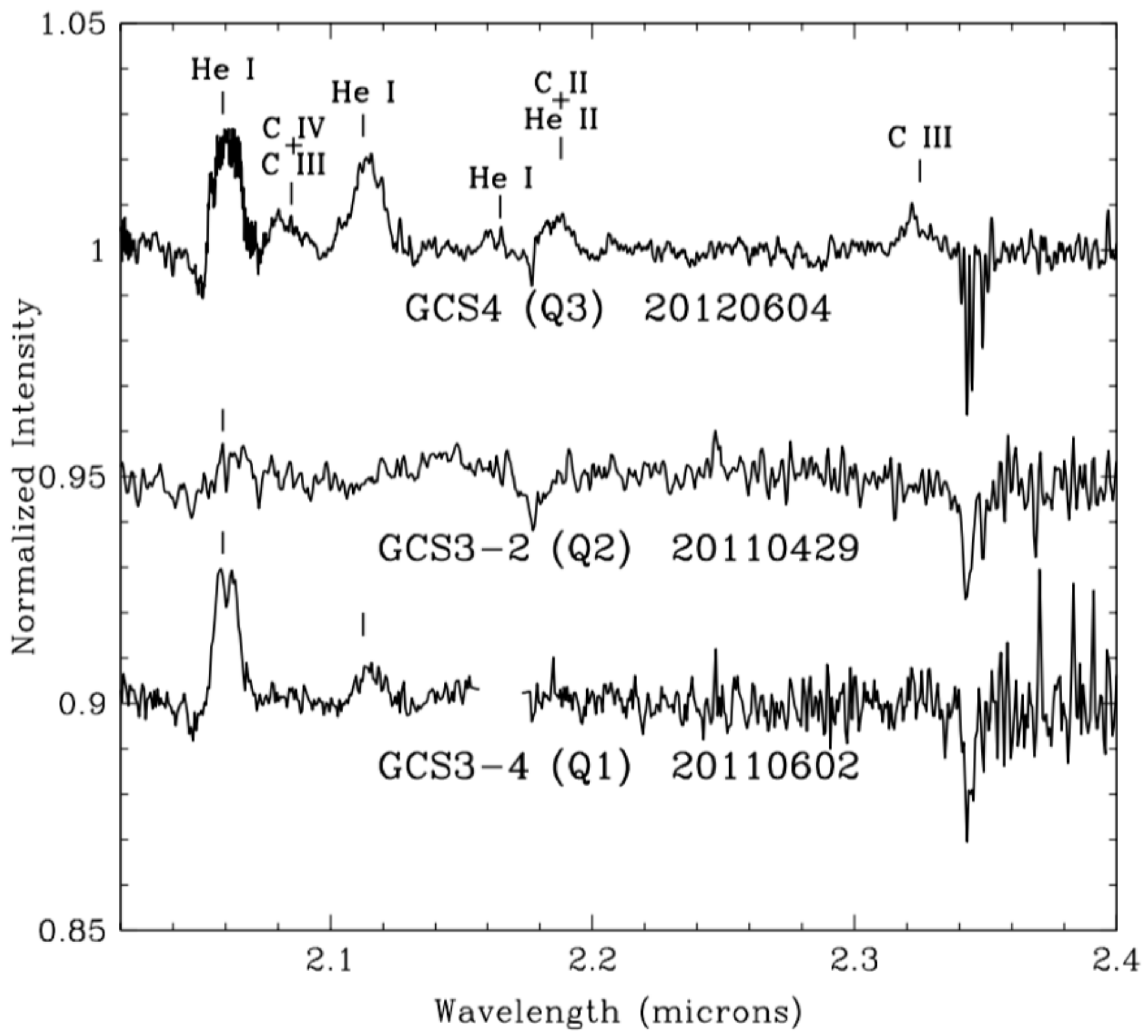}}
\end{minipage}
\caption{$K$-band continuum-normalized spectra of GCS4 ($R$=4000), GCS3-2 ($R$=1200), and GCS3-4 ($R$=1200).  Wavelengths and identifications of clearly detected emission lines are shown. Absorption near 2.35~$\mu$m is due to transitions between low lying rotational levels in the overtone band of CO in
foreground Galactic spiral arms. \label{fig:q_k}}
\end{figure}

Emission lines were detected in spectra of four of the five QPMs
(GCS3-1, GCS3-2, GCS3-4 and GCS4) as shown in Figs.~\ref{fig:q_j},
~\ref{fig:q_h}, and~\ref{fig:q_k}. For three of the QPMs lines were detected in the $J$ and $K$ bands, where previous lower sensitivity
observations found none \citep{fig99}. The $JHK$ spectra of GCS3-3 (Q9) were devoid of emission lines, perhaps due to a more opaque dust shell,
lower dust temperatures, and or variability, and are not shown. Line identifications (mainly from \citealt{een91} and \citealt{cro06}) are given
in Table~\ref{tab:lines}. In some cases the emission features may be blends of two or more lines.  Individual lines are well-resolved even at
the lowest resolving powers and have full widths at half maxumum of 1,500--3,000~km~s$^{-1}$. 
The singlet 2P-2S lines of \ion{He}{1} at 2.059~$\mu$m in both GCS4
and GCS3-4 have P Cygni profiles. Emission is present in the corresponding
triplet transition at 1.083~$\mu$m, but no P~Cygni absorptions were detected; however, the signal-to-noise ratios (S/Ns)
on the continua near that wavelength are very low.

Numerous absorption features are also present in the spectra of all five stars and are also listed in Table~\ref{tab:lines}. Most of these are already known diffuse interstellar bands (DIBs)
\citep{job90,geb11}, and apart from one DIB are in the $H$ band. Two DIBs, at 1.485~$\mu$m and 1.585~$\mu$m are reported here for the first time. The latter is easily seen in all five spectra in Fig.~\ref{fig:q_h}; the former is most clearly present in the spectra of GCS3-1 and GCS4. In the $K$ band, 
absorption lines due to overtone transitions of CO in low rotational levels are present near
2.35~$\mu$m (Fig.~\ref{fig:q_k}). These are known to arise in foreground spiral arms \citep{oka05}. 

Of the QPMs with emission lines, the richest spectrum in terms of number of detected lines is that of GCS4; the poorest is GCS3-2 (note, however, that there is no high sensitivity $J$-band spectrum of the latter). Carbon lines from the ionic species
\ion{C}{2} and \ion{C}{3} are generally prominent in GCS3-1, GCS3-4 and GCS4. Weak \ion{C}{4} line emission at 1.191~$\mu$m also can be seen in all three. Several \ion{He}{1} lines are also present in these stars. The \ion{He}{1} 1.083~$\mu$m has high equivalent widths in all three; longer wavelength \ion{He}{1} lines (at 1.278~$\mu$m  1.701~$\mu$m and 2.059~$\mu$m) have much lower equivalent widths because of their relative weaknesses, higher dilution by the continua of their dust cocoons, or both. \ion{He}{2} lines if present, are very weak and blended with \ion{C}{3} and \ion{C}{2} transitions, making their contributions uncertain. 

\begin{deluxetable}{lcc}
\tablewidth{0pt}
\tablecaption{Detected Lines and Possible Blends}
\tablehead{\colhead{Species} & \colhead{Wavelength (vac. $\mu$m)} & \colhead{Notes}}
\startdata
\ion{He}{1} & 1.083 & em: GCS3-1, 3-4, 4 \\ 
~\ion{C}{4} & 1.191 & em: GCS3-1, 3-4, 4 \\ 
~\ion{C}{3} & 1.198-1.199 & em: GCS3-1, 3-4, 4 \\
~\ion{C}{3} & 1.253-1.258  & em: GCS3-1, 3-4, 4 \\
~\ion{He}{1} & 1.278 & em: GCS3-4, 4 \\ 
~DIB & 1.318 & interstellar \\
~DIB & 1.485 & interstellar (new) \\
~DIB & 1.520 & interstellar \\
~DIB & 1.527 & interstellar \\
~DIBs & 1.56$-$1.57 & interstellar \\
~DIB & 1.584   & interstellar (new) \\
~DIB & 1.623 & interstellar \\
~DIBs & 1.657$-$1.660 & interstellar \\
~\ion{He}{1} & 1.701 & em: GCS3-1, 3-2, 3-4, 4 \\  
~DIBs & 1.77$-$1.80 & interstellar \\
~\ion{C}{2} & 1.784-1.786 & em: GCS3-1, 3-4, 4 \\
~\ion{He}{1} & 2.059  & em: GCS3-2, 3-4, 4 \\ 
~\ion{C}{4} & 2.071-2.080 & em: GCS4 \\ 
~\ion{C}{3} & 2.085 & em: GCS4 \\
~\ion{He}{1} & 2.113 & em: GCS3-4, 4 \\ 
~\ion{C}{3} & 2.114 & em: GCS4 \\
~\ion{He}{1} & 2.165 & em: GCS4 \\ 
~\ion{C}{2} & 2.188 & em: GCS4\\
~\ion{He}{2} & 2.189 & em: GCS4 \\
~\ion{C}{3} & 2.325 & em: GCS4 \\
~CO low $J$ & 2.34-2.36 & interstellar \\
\enddata
\label{tab:lines}
\tablecomments{All DIBs, except those at 1.485~$\mu$m and 1.585~$\mu$m 
have been reported previously \citep{job90,geb11}.}
\end{deluxetable}

\subsection{Spectral types}

The very broad emission lines observed in four of the QPMs are generally consistent with WR classifications. However, the superimposed dust emission in the spectra of the QPMs reduces the equivalent widths (EWs) of the detected lines by wavelength-dependent and widely varying amounts, and does not allow the typical analysis in which EWs of diagnostic lines in different wavebands are used to establish precise spectral types. To estimate the spectral types of the QPMs, we utilized line intensity ratios of transitions that are close in wavelength and are thus only slightly affected by assumptions about dilution by dust emission. For GCS3-1, GCS3-2, and GCS3-4 we used the \ion{C}{4} 1.191~$\mu$m to \ion{C}{3} 1.198--1.199~$\mu$m and the \ion{C}{4} 1.737~$\mu$m (not detected) to \ion{C}{2} 1.784--1.786~$\mu$m line ratios. For GCS4 we also utilized the \ion{C}{4} 2.071--2.084~$\mu$m to \ion{C}{3} 2.085~$\mu$m and 2.114~$\mu$m ratios (the latter contributing approximately two-thirds of the flux to the observed blend with \ion{He}{1} 2.113~$\mu$m).

We find that the spectra of the four QPMs resemble those of late-type WC stars \citep[e.g.][]{een91,cro06}. The relative weakness, or complete absence, of \ion{C}{4} at 1.191~$\mu$m and 1.736~$\mu$m in GCS3-1, GCS3-4, and GCS4, and of the latter line in GCS3-2, suggest that the spectral type of each is WC9 or later. All of these ratios are significantly less than one, similar to those of WC9 stars, and are much lower than the ratios for WC8 stars \citep[see Table 4 in][]{cro06}. Therefore, we assign the WC9d spectral type to each of the four QPMs with emission lines.

\subsection{Variability of GCS3-2 (Q2)}
Lines of \ion{He}{1} and \ion{C}{2} are present in the $H$-band spectrum of GCS3-2 obtained in 2010, but are absent in the 2011 $H$-band spectrum
(Fig.~\ref{fig:q_h}).  The singlet 2P--2S line of helium is weakly
present in 2011 (Fig.~\ref{fig:q_k}), but no $K$-band spectrum was obtained in 2010. \citet{tut06} have shown that GCS3-2 and GCS4 are dust-producing colliding wind binaries. As dust production varies with orbital phase, we suggest that the 2011 observations of GCS3-2 coincided with increased dust production, leading to stronger continuum emission as well as increased obscuration of the line-emitting region. 

\section{Modeling of CGS4 (Q3)}

\subsection{Previous models}

Previous characterizations of the QPMs \citep{fig99,mon01,tut06,don12,han16} have been hampered by the lack of detected stellar emission lines at near- and mid-infrared (NIR and MIR) wavelengths \citep{fig99,mon01}.  Only the properties of the dust shells, such as temperature and luminosity, and hints as to the properties of the central stars have been derived, by applying either blackbody/greybody fits to existing photometry \citep{fig99,don12}, or by more sophisticated dust models \citep{mon01,tut06,han16}, and by assuming an extinction law and guessing the contributions of stellar flux in the photometric wavebands. These assumptions, together with the spectral regions fitted, have yielded a range of possible dust temperatures and luminosities, but only vague constraints on stellar properties. 

For GCS4, \citet{fig99} obtained $T_{\rm shell~}$=~1055~K and
$\log(L_{\rm shell}/\Lsun)$~=~4.84 from fits to the NIR photometry using $A_K$~=~2.7~mag. The lack of line emission in their $J$-band spectra of the QPMs led them to suggest that the QPMs might be completely enshrouded in dust. \citet{mon01} fitted the 4-17~$\mu$m SED using $A_K$~=~3.3~mag and found $T_{\rm shell}$~=~700~K and $\log(L_{\rm shell}/\Lsun)$~=~4.6. Similarly to \citet{fig99} they concluded that for the $J$-band
spectra observed by those authors to be featureless, the ratio of dust shell luminosity to stellar luminosity would need to be higher than in the most extreme DWCLs known. On the other hand, \citet{don12} obtained $T_{\rm shell}$~=~644~K from NIR fits with $A_K$~=~3.4~mag, assuming equal flux contributions at 2.5~$\mu$m from the stellar source and the dust emission. Finally, \citet{han16} recently derived $T_{\rm shell}$~=~750~K and $\log(L_{\rm shell}/\Lsun)$=4.9 fitting the NIR (2MASS) and MIR (Spitzer/IRAC and SOFIA/FORCAST) photometry, and requiring $A_K$~=~1.8~mag, a  value that is in significant disagreement with the value of 3.1~$\pm$~0.5~mag found by \citet{lie10} for the Quintuplet Cluster.  

\subsection{Current model of GCS4}

The spectroscopy reported here provides, for the first time, the
possibility of utilizing the diluted spectra of the internal stellar
source(s), combined with the photometry, to much more fully constrain
these systems and characterize the embedded stars. We have proceeded
to model GCS4, the QPM for which the best data were obtained, assuming
that the system is described by a  WC~+~OB binary and a dust shell
\citep[see][]{cla11}. As in previous analyses of massive stars in the
Quintuplet Cluster \citep{naja09}, 
we adopt a distance of 8~kpc \citep{rei93,sch10,gil13}.

The normalized spectra  were modeled at each wavelength by adding the individual contributions to the total flux from the above three components and dividing their sum by the sum of their continua. We included the photometry out to 5~$\mu$m (well beyond the longest wavelength of these spectra). The full system is characterized by the total shell luminosity (out to that wavelength), the flux ratio of the summed stellar spectra to the dust shell at a given reference wavelength (we choose $\lambda=1.20~\mu$m), the flux ratio of the OB-star to the WC star at the reference wavelength, and the total (interstellar and circumstellar) extinction. For the latter we considered two different extinction laws. The first was derived by \citet{mon01} and is tailored to the Quintuplet Cluster. The second was obtained by \citet{nis06} using the red clump stars in the Galactic bulge. These extinction laws differ significantly in the NIR region, where the \citet{nis06} curve is much steeper. We found, however, that the two extinction laws yielded equally good fits if the $A_K$ values were adjusted independently, although the value of $A_K$, 2.54~mag, derived using the  \citet{nis06} law, may be unacceptably low, especially in that it includes  the contribution from the dust shell. Note also that the extinction law in the carbon-rich dusty shell is probably different from that of the interstellar dust in the 1--5~$\mu$m region, but that the shell provides only a small fraction of the total extinction, so the effect of using single extinction laws for both components is small. 

We adopted the NIR photometry of \citet{fig99} and the MIR SWS 4-17$\mu$m spectrum presented by \citet{mon01}, augmented with Spitzer/IRAC data. Given the photometric variability of these systems (\citealt{gla99} found peak-to-peak variations at $K$ of $\sim 0.44\pm0.06$~mag), the lack of simultaneous (spectro)-photometry over the modeled bands hinders an accurate determination of the luminosity. 

To constrain the WC star properties, we computed a grid of {\sc cmfgen} \citep{hil98} WC models and fitted the above described carbon line ratios which, together with the \ion{C}{3} 1.253-1.258~$\mu$m)/\ion{He}{1} 1.278 ratio, fixes the values of the stellar temperature (\Tstar) and mass loss rate (\Mdot) pairs as well as the C/He ratio. We also used the observed ratios of \ion{He}{1} and \ion{He}{2} lines in each waveband as a secondary ionization equilibrium constraint on the stellar temperature.  The P Cygni profile of the  \ion{He}{1}~2.059~$\mu$m line serves as a constraint on the stellar wind terminal velocity, \Vinf.

\begin{figure} 
\includegraphics[scale=0.57,angle=0]{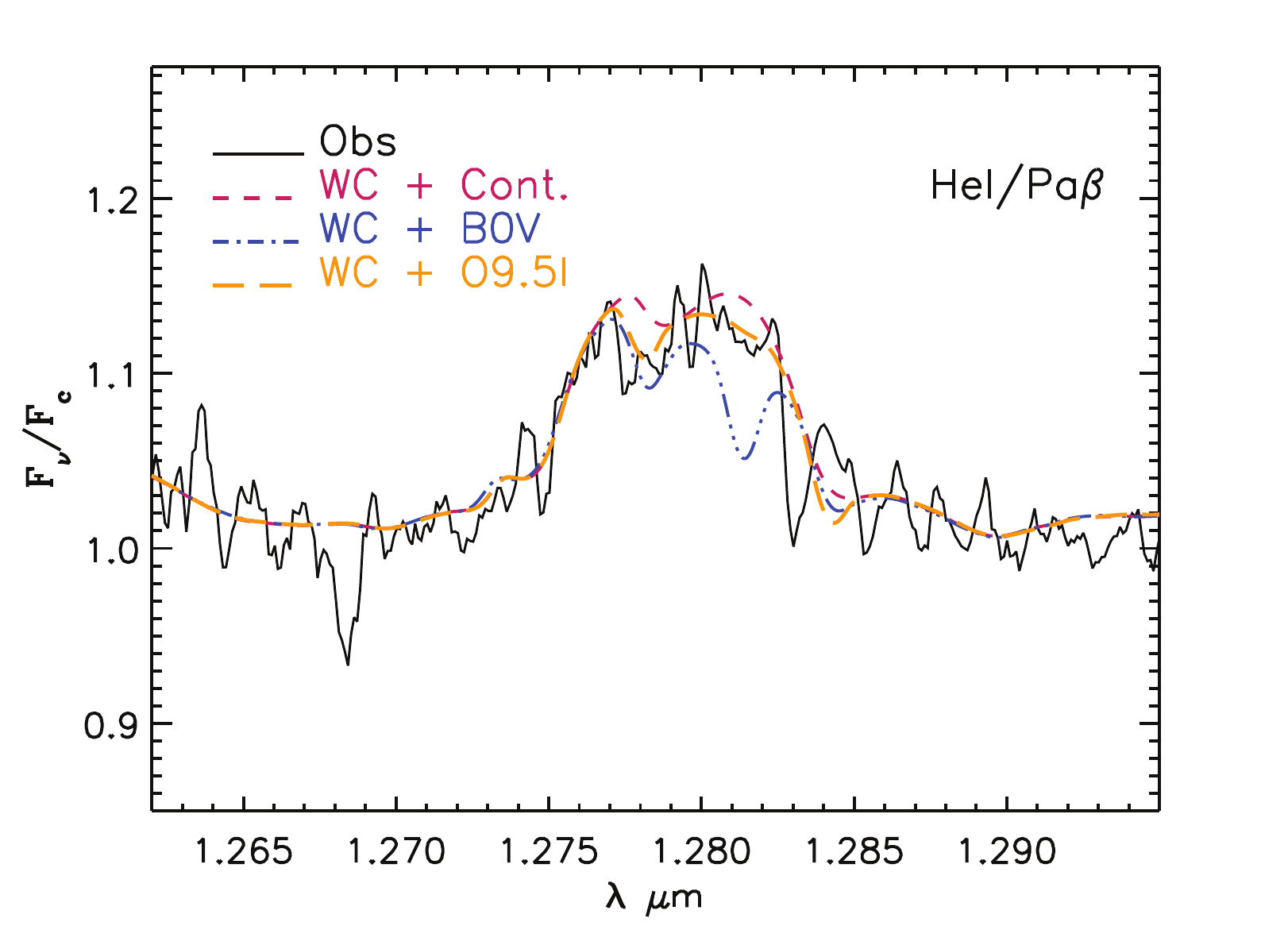}
\caption{Effect of the OB companion  on the observed Q3 (black-solid) \ion{He}{1}/Pa$\beta$ WC9 profile. Dashed (red) line
corresponds to the combination of the WC9 SED plus a hot featureless (only continuum) companion. Dash-dot (blue) and long-dashed (orange) display combinations of the WC9 stellar spectra with companions of spectral type B0V and O9.5Ia respectively. \ion{He}{1}/Pa$\beta$ is the best diagnostic for constraing the classification of the OB companion.
\label{fig:gcs4-pabeta}}
\end{figure}

To characterize the OB component of the system, we selected models from our own {\sc cmfgen} grid of OB stars to investigate
how stars of different spectral types from early O to early B and from dwarfs to supergiants could leave their spectral imprints on the observed profiles of the WC star.  If present, we expect  the hydrogen and/or helium lines of the OB companion
to appear  in the form of relatively narrow absorption dips on top of the broad emission lines of the WC star.  
From the point of view of the overall system, shorter wavelengths should be more optimal for detecting the signature of an OB star, as the relative contribution from
that star will be larger and hence its spectral lines will be less diluted. However, the strong extinction dramatically lowers the S/N below 1.20~$\micron$ and hampers the detection of the OB features. On the other hand, at longer wavelengths ($\lambda \geq 2.0\micron$) where a high S/N is achieved, the greater contribution of dust and the increasing WC/O flux ratio (due to wind bound-free and free-free emission of the WC star) drastically attenuate the OB line strengths. Thus, within our data set, we found that the best OB diagnostic
is provided by the \ion{He}{1}/Pa$\beta$ 1.28~$\micron$ feature and, to a lesser extent, by the \ion{He}{1}~$1.701~\micron$ line and \ion{H}{1} lines in the $H$ band. This is shown at 1.28~$\mu$m in the model spectra in Fig.~\ref{fig:gcs4-pabeta}, where the imprints of a late O supergiant and a B0V star are clearly visible on top of the broad \ion{He}{1}/Pa$\beta$ emission line from the WC star.

Once the main properties of the WC+OB system were derived, we computed a set of dust shell SEDs with different temperatures to fit the photometric data out to 5~$\mu$m. We first used the approach of \citet{don12} and \citet{mon01}, assuming a diluted blackbody with a $\lambda^{-1}$ emissivity. However,  we found \citep[see also][]{mon01} that this characterization produced SEDs that are too narrow when compared to the observations. We then opted for simple blackbody emission, which yielded broader SEDs and quite similar results to those obtained from more complex dust emission models  \citep[e.g.][]{mon01}.

The selected WC, OB star and dust shell models were then scaled to reproduce the diluted $J$-, $H$- and $K$-band profiles.
The total fluxes were combined into a single SED, which was reddened and rescaled to fit the observed photometry. The total
luminosity of each component was then determined.

\section{Results of Modeling and Discussion}

\subsection{Derived physical properties}

\begin{figure} 
\centering
\begin{minipage}{4in}
{\includegraphics[scale=0.375,angle=90]{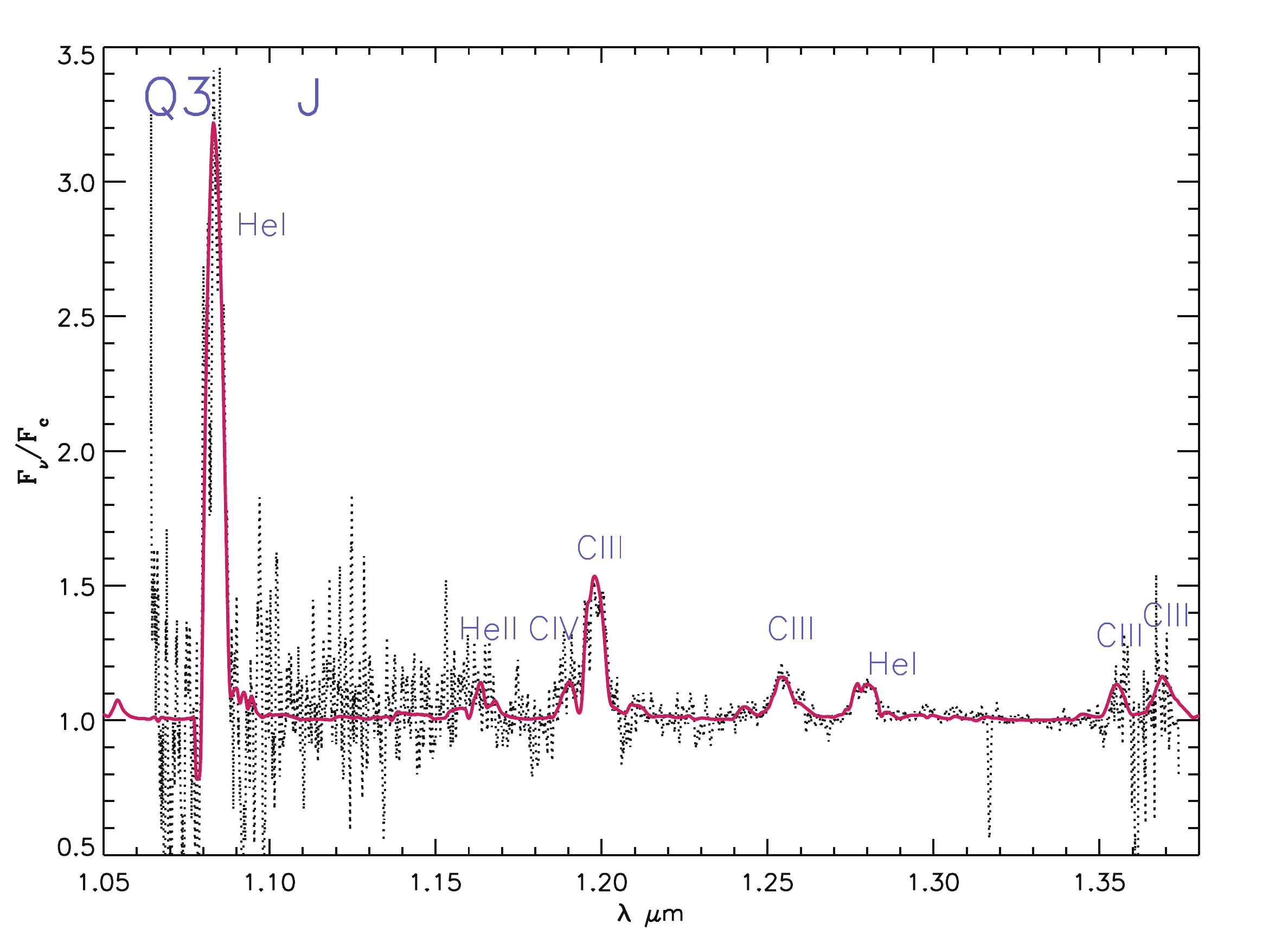}}
\end{minipage}
\\
\vspace{-2mm}
\begin{minipage}{4in}
{\includegraphics[scale=0.375,angle=90]{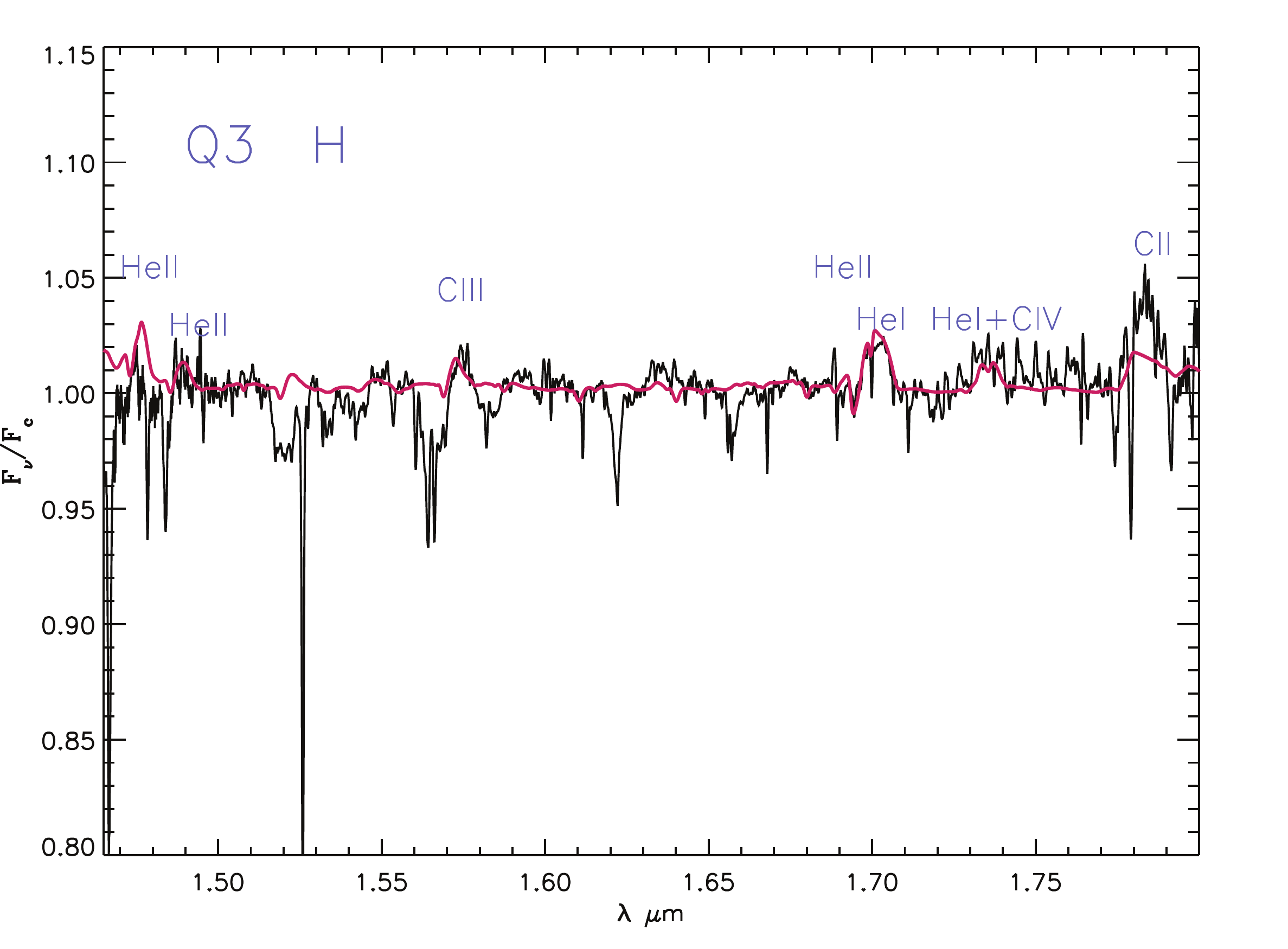}}
\end{minipage}
\\
\vspace{-2mm}
\begin{minipage}{4in}
{\includegraphics[scale=0.375,angle=90]{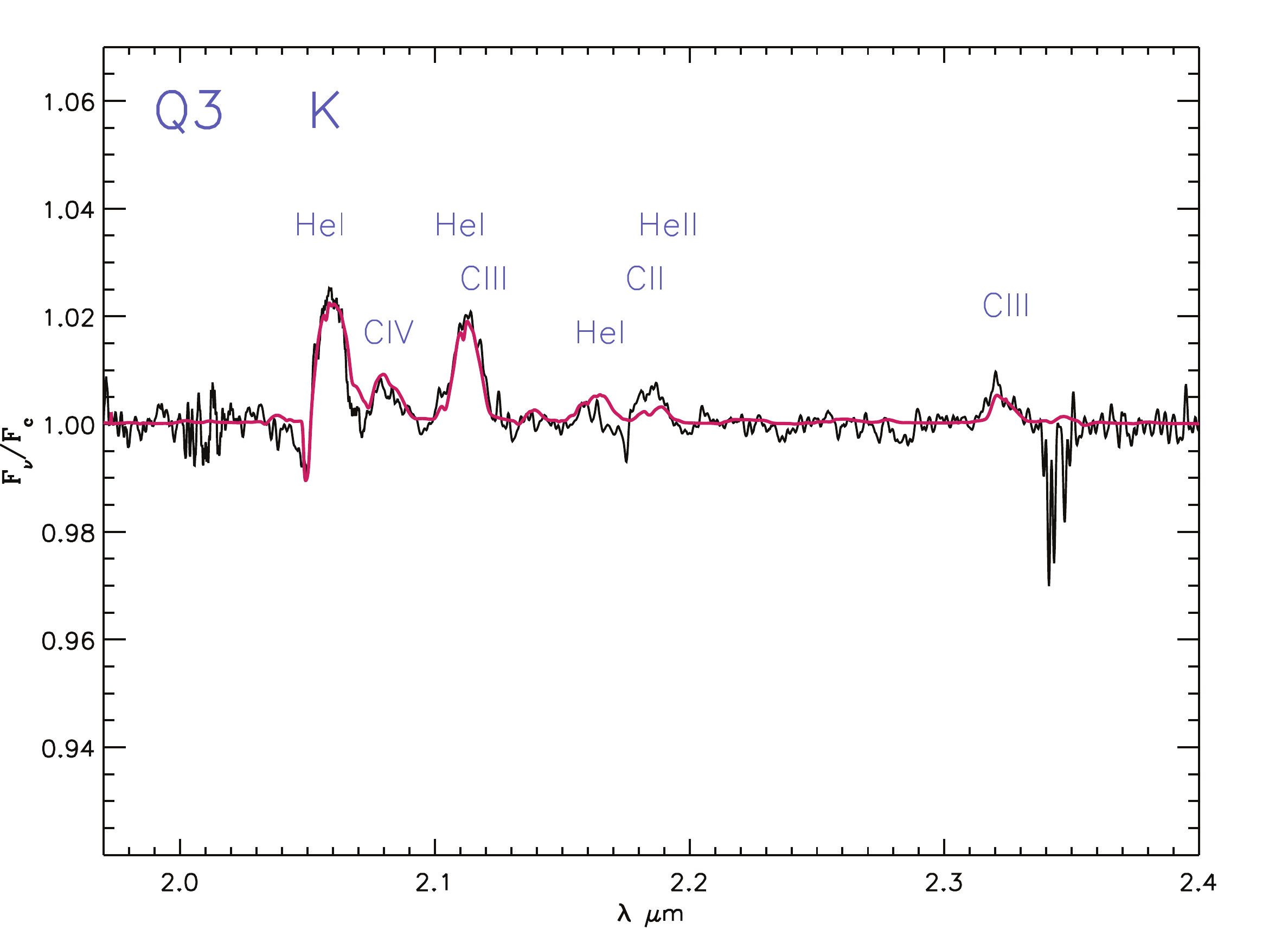}}
\end{minipage}
\caption{Comparison of the normalized spectra (dotted lines) of GCS4(Q3) with that for a theoretical model (solid lines) 
consisting of contributions from a WC9 star, an O-star, and dust (see
text).\label{fig:gcs4_jhkmodel}}
\end{figure}

\begin{figure} 
\includegraphics[scale=0.385,angle=90]{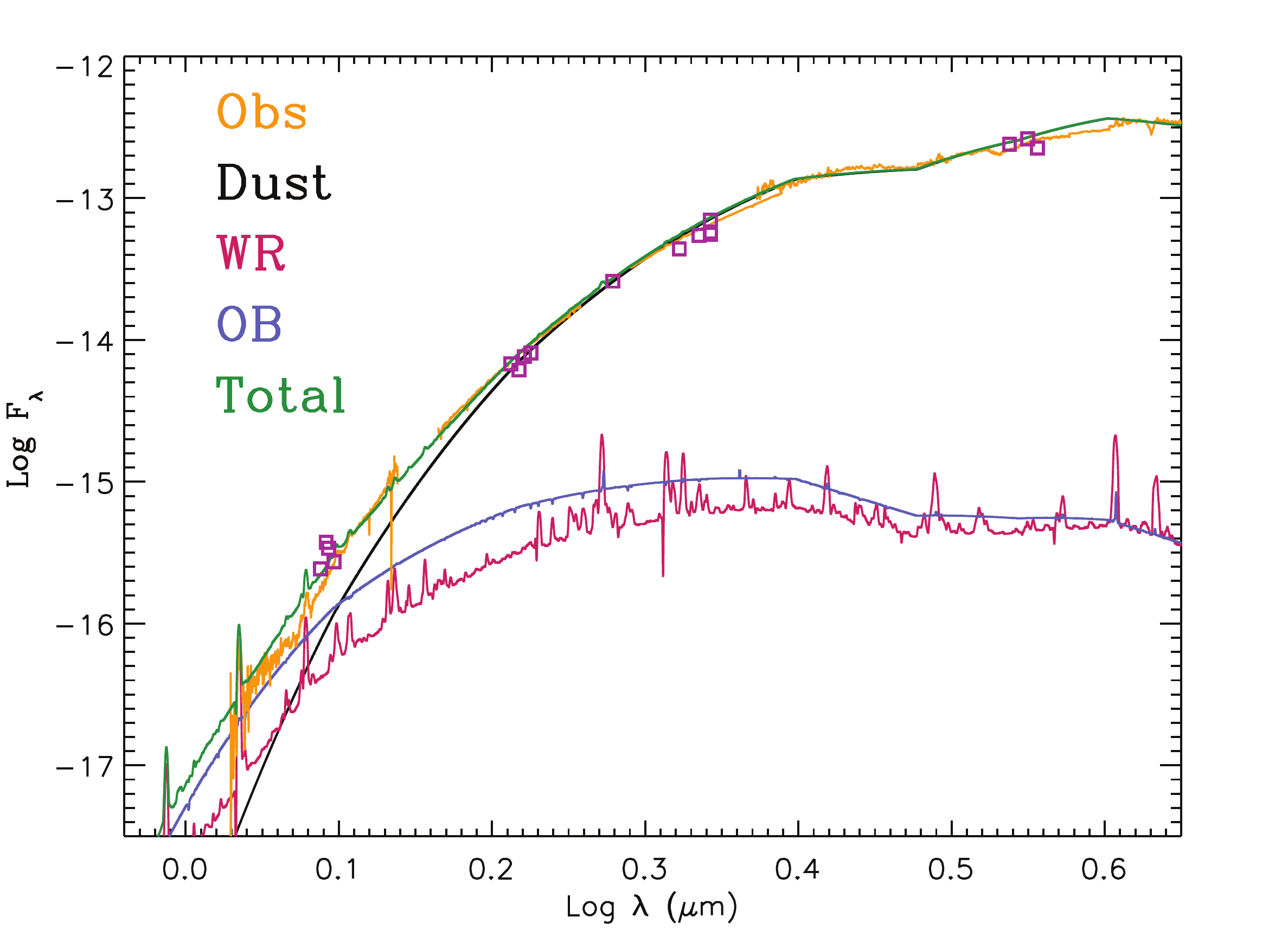}
\caption{Fit to the observed SED of GSC4 (orange),
including data from from ISO/SWS. 
We assume that the total flux (green) is due to the
  contribution by dust (black), the WC9d WR star (red) and an OB
   star (blue). Photometry is symbolized by squares.
\label{fig:gcs4-full}}
\end{figure}

Table~\ref{tab:params} lists the derived properties of GCS4.  Those that depend on the assumed extinction law are displayed separately for each of the two laws. Figure~\ref{fig:gcs4_jhkmodel} shows the excellent fits of the models to the observed $J$-, $H$-, and $K$-band diluted spectra while Fig.~\ref{fig:gcs4-full} displays the comparison between the observed spectrophotometry and the SED of the reddened  WC+OB+dust~shell. The individual contribution of each component is also plotted. From Fig.~\ref{fig:gcs4-full} it is clear that the dust shell dominates the total flux at wavelengths as short as the $H$ band, while its contribution in the $J$ band is similar to the WC+OB system. At the reference wavelength, 1.20~$\mu$m, we derive F$_{\rm shell}$/F$_{\rm WC}$=1.6 and F$_{\rm shell}$/F$_{\rm OB}$=0.6. \footnote{We also found that the dust emission can be fit well by a power law relation, i.e., F$_\lambda \propto \lambda^6$. If we further assume that three-fourths of the flux at 1.2~$\mu$m is attributable to this thermal emission, the resulting line strengths and ratios are also matched.}

Almost all of the strong  carbon and helium emission lines are well
reproduced by the model (see Fig.~\ref{fig:gcs4_jhkmodel}),
the only notable exception being the \ion{C}{2} feature at 1.784-1.786~$\mu$m,
which is commonly underestimated by the models, \citep[e.g.,][]{cla11}.
This doublet is very sensitive to blanketing as its upper levels are
fed through the 2s$^2$2p$^2$P-2s$^2$4d$^2$D resonance lines at
$595\AA$. There are a few other more minor discrepancies.
The slight mismatch in the \ion{He}{2} 1.476~$\mu$m line is caused by a
combination of the DIBs and the position of the line at the blue-edge
of the H-band which hinders a proper normalization.
The small discrepancy present at \ion{He}{1}~2.164$\mu$m could be due
to the residuals from the  Br$_{\gamma}$ removal of the standard star, while the
weaker $2.189\mu$m feature produced by the models (a blend of the 
\ion{He}{2}~$2.189\mu$m line with the \ion{C}{2}~$2.188\mu$m n=10 to n=7 transitions)
is probably caused by the models underestimating the latter. 

The ratio of the close lying \ion{C}{4} and \ion{C}{3} lines in the $J$ and $K$ bands fixes the pair values given by the WC stellar temperature T$_{*}$ (defined at a Rosseland opacity, $\tau_{Ross}$ of 10) and the so-called ``transformed radius'' \Rt$= R_\star [ (v_\infty/2500) / (10^{4}\Mdot/\sqrt{f})]^{2/3}$ \citep[where $R_\star$ is in solar radii, \Vinf\ is in \kms\ , \Mdot\ is in \Msunyr, and $f$ is the clumping value;][]{sch89}.

We derive T$_{*}$~=~49.3~$\pm~2.5$~kK and \Rt~=~4.2 for the WC star. We find that \Vinf\ = 1250~$\pm~150$~km~s$^{-1}$ yields an excellent fit to the \ion{He}{1} P Cygni-profiles at 1.701~$\mu$m and 2.059~$\mu$m and satisfactorily matches the observed widths of the emission lines as well as the shapes of the line blends. Our value is roughly 30\% lower than the 1800~$\pm~300$~km$^{s}$ suggested by \cite{tut06}, who fit an Archimedian spiral model. Depending on the extinction law applied (\citeauthor{mon01} or \citeauthor{nis06}), we obtain different values of \Rstar\, yielding respective values of \Lstar\ and \Mdot\ (we assume a clumping value, $f$=0.1) as displayed in Table~\ref{tab:params}. Finally, we derive a mixture of 20\% carbon and 78\% helium (mass fraction) at the stellar surface. 

The OB companion can only be poorly constrained. Making use of the observed shape of the \ion{He}{1}/Pa$\beta$~1.28-$\micron$ complex (see Fig.~\ref{fig:gcs4-pabeta}) and noting that the OB star is roughly 2.5 times brighter than the WC in the continuum at this wavelength, we may discard a mid-O to mid-B dwarf companion as its Pa~$\beta$~1.28-$\micron$ line would be clearly detectable. Hence, the presence of a supergiant with a strong enough wind to significantly fill the absorption profile is favored.  Further, if we assume that the dip at 1.277$\mu$m is due to \ion{He}{1}, then the supergiant cannot be too hot. We find that stars with spectral types between O8If and B0Ia can match the observed \ion{He}{1}/Pa$\beta$~1.28-$\micron$ blend and that an excellent fit is obtained for an O9.5Ia star such as $\alpha$~Cam \citep{naja11}.
The lack of precise information on the properties of the binary system prevents us from applying any phased radial velocity correction to the OB companion in Fig.~\ref{fig:gcs4-pabeta}. Nevertheless, given the low inclination of the system \citep{tut06} we expect a rather small projected radial velocity shift from the orbital motion.

\begin{deluxetable}{llr}
\tablewidth{0pt}
\tablecaption{GCS4 derived properties}
\tablehead{\colhead{Extinction law} & \colhead{Property} & \colhead{Value} }
\startdata
              & T$_{\rm shell}$ (K) & 1000\\
              & T$_{* \rm WC}$ (kK)$^a$ & 49.3\\
              & XC$_{\rm WC}$$^b$    & 0.20\\
              & \Vinf$_{\rm WC}$ (\kms)  & 1250\\
              & L$_{\rm shell}$/\Lstar$^c$ & 0.17\\
\citet{mon01}: &                     &     \\ 
              & $A_K$            & 3.30\\
              & log\,L$_{\rm shell}$/\Lsun  & 5.22\\
              & log\,L$_{\rm WC}$/\Lsun  & 5.44\\
              & log\,L$_{\rm OB}$/\Lsun$^d$  & 5.84\\
              & R$_{* \rm WC}$/\Rsun$^a$ & 7.15\\
              & Log (\Mdot/$\sqrt{f}$) (\Msunyr)$^e$  & -3.95\\
\citet{nis06}: &  & \\ 
              & $A_K$ & 2.54\\
              & log\,L$_{\rm shell}$/\Lsun  & 4.87\\
              & log\,L$_{\rm WC}$/\Lsun  & 5.09\\
              & log\,L$_{\rm OB}$/\Lsun$^d$  & 5.49\\
              & R$_{* \rm WC}$/\Rsun$^a$ & 4.78\\
              & log (\Mdot/$\sqrt{f}$) (\Msunyr)$^e$  & -4.21\\
\enddata
\label{tab:params}
\tablecomments{a:\,Values at $\tau_{Ross}$=10. b:\,Mass fraction. 
c:\,WC+OB luminosity. d:\,Assuming a O9.5Ia (see text). e:\, Models
assume a clumping value {\it f}=0.1, so the true mass-loss rates will be
0.5~dex lower.}
\end{deluxetable}

The derived temperature of  $T$~=~1000~$\pm$~100~K  for the dust shell agrees very well with the one obtained by \citet{fig99} from NIR fitting and differs considerably from the 640-750~K values derived by \citet{mon01}, \citet{don12}, and \citet{han16}. The  combination of the SED and the diluted spectrum is fundamental to estimating the uncertainty in $T_{\rm shell}$. For high dust temperatures, e.g. $T_{\rm shell}$~$>$~1100~K, an acceptable fit to the $JHK$ photometry is still possible by increasing the interstellar extinction and doubling the flux ratio of the shell to the WC star, F$_{\rm shell}$/F$_{\rm WC}$.
However, the model then clearly underestimates the observed flux longwards of 2.4~$\mu$m and the spectral lines in the shorter wavelength portion of the $J$ band become unacceptably strong. Imposing a lower dust temperature, e.g. $T_{\rm shell}$~$\sim$~700~K, as derived by other studies makes it impossible to simultaneously fit the NIR SED and the normalized diluted spectra. Besides, an unacceptably large vale of  F$_{\rm OB}$/F$_{\rm WC}$ ($\sim$~4) is required, and our models show that the stellar features from the OB companion would have higher EWs in the $H$ band than are observed (see below). Moreover, although a reasonable fit to the SED is obtained up to $\lambda \approx$~20$~\mu$m, a very low value of $A_K$ is obtained and the $J$-band photometry is overestimated by roughly two magnitudes independent of the extinction law used.

Although the model in Table~\ref{tab:params} is a very good fit for
$\lambda < 6~\mu$m, it underestimates the observed SED at longer
wavelengths. A better fit at $\lambda>6~\mu$m, would  require the
presence of a cooler component, which would have only a modest effect on the values in the table. Indeed, the use of a single temperature for the dust shell modeling for the Quintuplet
members was questioned by \citet{tut06}, who resolved GCS4  at 2.21 and 3.08~$\mu$m.  A larger increase in size between 2.21 and 3.08 $\mu$m  (factor of $\sim$2) was found for this object than for other pinwheel systems in the Galaxy  \citep[factor of $\sim$1.4, ][]{mon07} and was attributed to a range of temperatures contributing to the 2--3-$\mu$m emission, which would cause a flatter thermal profile from the GCS4 dust shell.

\subsection{Comparison with previous studies}
 
Here we discuss the discrepancies between our results for GCS4 and those obtained by \citet{mon01}, \citet{don12} and \citet{han16}. The model fits, with $T_{\rm shell}$~$\sim$~700~K presented by \citet{mon01} (their Fig. 11), tailored to the 4--17~$\mu$m observations, clearly underestimate the NIR bands and imply the need for a second, hotter dust component.  \citet{don12} fit model SEDs to the NIR photometry for all five of the QPMs. They assumed, in each case, that the $J$-band flux is heavily dominated by flux from the star and that the $K$-band flux is roughly split evenly between contributions from the star and the dust (see their Figure~4). This assumption is untenable as it would produce high equivalent width emission lines from the WC star in the $K$ band, as the stellar lines would then be diluted just by a factor of $\sim$2 (compared to the factor of $\sim$~100 observed in GCS4). 

\citet{han16}, who reported mid-infrared photometry of the QPMs, extended their derived SEDs to the near-infrared data using DUSTY models, assuming a spherical  dust shell illuminated by a central source, which they chose to be a 40,000~K blackbody with a Rayleigh-Jeans like behavior in the NIR. They preferred a 750$\pm$50\ K blackbody for GCS4, noting the good fit between such a model and the data, although the fit at the shortest wavelengths in the near-infrared is not as good as at longer wavelengths. However, their fit requires  $A_K$~=~1.8~mag for the interstellar extinction, which is roughly 0.6~mag below the lowest estimates presented in the literature for the Quintuplet cluster. We adopted the above temperatures and extinction derived by \citet{han16}
and obtained a similar quality fit to the full 1.25--37~$\mu$m photometry. However, we failed to reproduce the observed normalized spectra. At short wavelengths the model severely overestimates the observed line strengths while the lines in the model $K$-band spectra are much too weak when compared to the observations. Although the disagreement could be considerably reduced by invoking a second hotter dust component, the mismatch again emphasizes the necessity of combining the observed diluted spectra with the photometry to more tightly constrain the properties of the QPMs.

\section{The QPMs vs related Pinwheel and  WC9d objects.}

Our derived properties for GCS4 can be compared to those of well studied Pinwheel systems such as WR104 \citep{tut99,har04,tut08}. 
The latter is composed of a WC9d + B0.5V binary \citep{cro97} with a period of 241~days \citep{tut08}. 
The binary system was modeled by \citet{cro97} assuming a flux ratio OB:WC~=~2 in the $V$ band with the WC9 having almost identical properties
(T$_{*}=45$~kK, \Rt=4.0 and \Vinf=1220~\kms) to the ones we have derived here for GCS4 (see Table~\ref{tab:params}). Interestingly, our derived value for the dust temperature in GCS4 (T=1000~K) agrees very well with the azimuthally-averaged temperature in the inner regions as inferred from the model by \citet[][their Fig.~5]{har04}. The systems' luminosities differ significantly (log~$\Lstar_{WR104}/\Lsun = 4.6$), assuming the distance of 1.6~kpc adopted
by \cite{cro97} and by \cite{har04}. However, this value was revised up to 2.6~kpc by \cite{tut08} after combining the measured physical speed of the dust plume with the apparent angular velocity. The revised distance results in log~L/\Lsun=5.5 for the binary system \citep[e.g.][]{mon07}, which agrees very well with the one we obtain for GCS4 assuming the \cite{nis06} extinction law, and emphasizes the importance of an accurate knowledge of the distance as well as a good estimate of the extinction for these objects. 
Indeed the lack of reliable distance estimates for Galactic WC9 stars has
led to some studies assigning an absolute magnitude $M_{v}$ for these
stars. Some have adopted $M_{v}=-4.6$ \citep[as suggested
by][]{van01}, while brighter magnitudes ($M_{v}=-5.17$) have been used
for the WC9 star in the systematic study of WC Galactic stars by
\cite{san12}, yielding average luminosities of $log$~L/\Lsun~=~5.2 for
the WC9d stars.\footnote{We expect Gaia DR2 to provide a definite
M$_v$ calibration.} Interestingly, this value lies in between our two
estimates of the luminosity of the GCS4 WC9 star (Table~\ref{tab:params}). We may as well use luminosity calibrations  as a function of spectral type \citep[e.g.,][]{mar05} for the OB companion to discern the luminosity class of the latter and/or the appropriate extinction law. Such a method may prove efficient for systems with both a solid determination of the companion's spectral type and a relatively low observed luminosity scatter (e.g. late O and early B dwarfs). However, in the case of WR104, where a B0.5V companion has been assumed \citep{cro97,har04}, its inferred value of log~L/\Lsun~5.3  (at 2.6kpc) would imply a drastic revision of its spectral type and/or luminosity class  (e.g. either a O6V or a B0.5I would be required). In the case of GCS4, the large scatter in luminosities observed within the O9.5Ia spectral types does not allow us to differentiate between the two extinction laws. The value log~L/\Lsun=5.84 for the \cite{mon01} law coincides with the luminosity of $\alpha$~Cam, one of the canonical O95.Ia stars \citep{naja11} while the log~L/\Lsun~5.49 solution for the \cite{nis06} extinction law also matches perfectly the \cite{mar05} calibrated luminosity value for a O9.5I star. 

The SEDs of the QPMs closely resemble that of W239, a dusty WC star in Westerlund 1 \citep{cro06,cla11}. Despite W239 having a much shorter period $\sim5$~days vs $\sim850$~days for GCS4 \citep{gla99}, the spectral similarities between the two systems are striking.  \citet{cla11} fit the SED of the W239 system (their Figure~5) with contributions from a WC+O binary plus a 1300~K greybody (slightly hotter than GCS4). Assuming that the O dwarf companion and the WC9 star have equal fluxes at $8500\AA$, they obtain a $J$-band flux ratio of 2:1 between the binary and the dust. While this ratio is similar to the one we derive for GCS4, the relative contribution of each of the binary components differs considerably, with the O9.5Ia contributing $\sim 70\%$ of the $J$-band binary flux in GCS4 and the W239 O dwarf with $\sim 50\%$. This result is consistent with the luminosity classes derived for each of the systems. 

The different dust temperatures and WC9/OB flux ratios of the two 
systems are nicely traced by the observed dilution of the spectral features
in the different bands. While both systems show nearly identical equivalent widths in the $J$-band, e.g., 35~\AA\ for the \ion{C}{3} line at 1.20~$\mu$m for  GCS4 and W239, in the $H$ band, the spectral features of the QPMs are more diluted. This is clearly tracked by the equivalent widths for the \ion{He}{1} 1.701$\mu$m lines (2~\AA\ and 5~\AA\ for GCS4 and W239, respectively), suggesting that the dust temperature must be lower for GCS4 than for W239, as found by our modeling (see Table~\ref{tab:params}). As confirmation, note that the \ion{He}{1} 2.11~$\mu$m line peak in the $K$-band spectrum of W239 in \citet{cla11}(Fig.~6, bottom) is  $\sim3.5$ higher above the continuum than in GCS4 (see Fig.\ref{fig:gcs4_jhkmodel}) and is basically absent for the rest of the QPMs, again suggesting a very low ratio of light from the stellar system to that from the dust.
 The stellar properties of the WC9 component in W239 are very close to
those derived for GCS4 when applying  the \cite{nis06} extinction law. Although these results seem
to favor the extinction law  of \citet{nis06} vs. that of
\citet{mon01}, we cannot exclude the latter until the OB component, and hence the full system, is better constrained.

\section{Conclusion}

While the QPMs were first noted for their remarkable infrared SEDs, their characteristics, and presence in the Quintuplet cluster, are easy to understand when considering that they are members of a massive cluster of stars at an age when some of the O-stars have evolved away from the main sequence. They have average properties for dusty WC stars, including their emission lines, colors, and brightnesses. Their inferred luminosities, temperatures, dust shells, and binary natures also are typical of dusty WC9 stars. Combined with previous observations of pinwheels that surround two of the stars \citep{tut06}, we conclude that all five of the QPMs are dusty WC9 stars, as initially suggested by \citet{fig99}.

\begin{acknowledgements}

The data presented here were obtained at the Gemini Observatory, which is operated by the Association of Universities for Research in Astronomy, Inc., under a cooperative agreement with the NSF on behalf of the Gemini partnership: the National Science Foundation (United States), the National Research Council (Canada), CONICYT (Chile), Ministerio de Ciencia, Tecnolog'a e Innovaci—n Productiva (Argentina), and MinistŽrio da Cincia, Tecnologia e Inova‹o (Brazil).
We thank John Hillier for providing the {\sc cmfgen} code. F.N. and D.dF. acknowledge grants FIS2012-39162-C06-01, ESP2013-47809-C3-1-R and ESP2015-65597-C4-1-R

\end{acknowledgements}

\end{document}